\documentclass[preprint2]{emulateapj-rtx4}

\newcommand{\kms}{{\km\second^{-1}}}

\newcommand{\km}{\,{\textnormal{km}}}
\newcommand{\second}{\,{\textnormal{s}}}

\newcommand{\dechms}[4]{$#1^{\rm h}#2^{\rm m}#3\mbox{$^{\rm s}\mskip-7.6mu.\,$}#4$}

\newcommand{\decdms}[4]{$-#1^{\circ}#2'#3\mbox{$''\mskip-7.6mu.\,$}#4$}

\usepackage{txfonts} \usepackage{epsfig} \usepackage{graphicx}
\usepackage{natbib} 
\begin{document}

\title{ALMA OBSERVATIONS OF THE OUTFLOW FROM THE SOURCE I IN THE ORION-KL REGION}

\shortauthors{Zapata, et al.}

\author{Luis A. Zapata\altaffilmark{1}, Luis F. Rodr\'\i guez\altaffilmark{1}, Johannes Schmid-Burgk\altaffilmark{2},
 Laurent Loinard\altaffilmark{1,2}, 
Karl M. Menten\altaffilmark{2}, and Salvador Curiel\altaffilmark{3}} 


\altaffiltext{1}{Centro de Radioastronom\'\i a y Astrof\'\i sica, UNAM, Apdo. Postal 3-72
  (Xangari), 58089 Morelia, Michoac\'an, M\'exico}
\altaffiltext{2}{Max-Planck-Institut f\"{u}r Radioastronomie, Auf dem H\"ugel
  69, 53121, Bonn, Germany} 
\altaffiltext{3}{Instituto de Astronom\'\i a, Universidad Nacional Aut\'onoma de M\'exico, Ap. 70-264, 04510 DF, M\'exico} 

\begin{abstract} 
  
 In this {\it Letter}, we present sensitive millimeter SiO (J=5-4; $\nu$=0) line observations of the 
outflow arising from the enigmatic object Orion Source I made with the Atacama Large 
Millimeter/Submillimeter Array (ALMA). The observations reveal that at scales of a few thousand AU, 
the outflow has a marked "butterfly'' morphology along a northeast-southwest axis. However, contrary 
to what is found in the SiO and H$_2$O maser observations at scales of tens of AU, 
the blueshifted radial velocities of the moving gas are found to the northwest, 
while the redshifted velocities are in the southeast. The ALMA observations are  
complemented with SiO (J=8-7; $\nu$=0) maps (with a similar spatial resolution) obtained 
with the Submillimeter Array (SMA). These observations also show a similar morphology 
and velocity structure in this outflow. We discuss some possibilities to explain 
these differences at small and large scales across the flow.   
\end{abstract}


\section{Introduction}

Since its discovery, the radio Source I in the middle of the Kleinmann-Low nebula in Orion, has always been 
an enigmatic object \citep{men1995,chu1987}.
Recent proper motion measurements suggest that the Orion Source I might have been part of a multiple system 
that disintegrated some 500 years ago \citep{gom2008,rod2005,god2011} originating an explosive outflow
in the Orion BN/KL region \citep{zap2009,bally2011,zapata2011a}. \citet{bally2011} proposed that probably the
Orion Source I is a short-period binary formed just before the stellar ejection and the formation 
of the explosive outflow. It is estimated that the separation of the binary is only a few Astronomical Units (AU) 
\citep{bally2011}.

Strong H$_2$O and SiO vibrationally excited ($v = 1$ and 2) masers are concentrated very close to the Orion Source I. 
The H$_2$O masers are distributed 
in an elongated pattern along a northeast-southwest axis and seem to be expanding about Source I \citep{gen1981}. 
The bulk of the SiO masers are found surrounding Source I and form an X-shaped structure; 
the emission in the southeast arms is predominantly blueshifted, while the emission in the northwest is predominantly 
redshifted. The intermediate-velocity emission is observed connecting the red 
and blue sides of the emission distribution. The SiO masers seem to be arising from a wide-angle bipolar 
wind emanating from a rotating, ionized edge-on disk with a southeast-northwest orientation \citep{matt2010}. 
The ionized disk has been imaged at 7 mm. and appears to be surrounding a young massive star \citep{men2007,god2011}. 

At scales of a few hundreds of AU, \citet{plam2009} imaged the outflow from the Orion Source I using 
CARMA (The Combined Array for Research in Millimeter-wave Astronomy) in the J = $2-1$ from the vibrational ground state of SiO, 
and found that the central velocity channels showed the limb-brightened edges of two cones centered on Source I. 
However, since the outflow appears to lie nearly in the plane of the sky, 
there is no measurable red-blue velocity asymmetry in the cones, as observed for the masers.  

We report $\sim$ 1.5$''$ resolution images in the SiO J=5-4; $\nu$=0 
obtained with the Atacama Large Millimeter/Submillimeter Array\footnote{ALMA is a partnership of ESO (representing its member states), 
NSF (USA) and NINS (Japan), together with NRC (Canada) and NSC and ASIAA (Taiwan), 
in cooperation with the Republic of Chile. The Joint ALMA Observatory is operated by ESO, AUI/NRAO and NAOJ.} (ALMA)
from the Orion Source I. These observations are  
complemented with SiO (J=8-7; $\nu$=0) maps obtained 
with the Submillimeter Array\footnote{The Submillimeter Array (SMA) is a joint project between the Smithsonian
Astrophysical Observatory and the Academia Sinica Institute of Astronomy and 
Astrophysics, and is funded by the Smithsonian Institution and the Academia Sinica.} (SMA). 
Both observations reveal that the outflow from the Orion Source I has a \textquotedblleft butterfly'' morphology 
along a northeast-southwest axis. The blueshifted radial velocities are found to the northwest, 
while the redshifted velocities toward the southeast. 

\section{Observations}

\subsection{ALMA}

The observations were made with sixteen antennas of ALMA on January 2012, during the
ALMA science verification data program. 
The array at that point only included antennas with diameters of 12 meters.
The 120 independent baselines ranged in projected length from 18 to 253 k$\lambda$. 
The phase reference center for the observations was at $\alpha_{J2000.0}$ = 
\dechms{05}{35}{14}{35}, $\delta_{J2000.0}$ = \decdms{05}{22}{35}{00}.
The primary beam of ALMA at 217 GHz has a FWHM $\sim 30''$. The emission from the outflow powered by 
the Orion Source I falls very well inside of the FWHM.

\begin{figure}[ht]
\begin{center}
\includegraphics[scale=0.45]{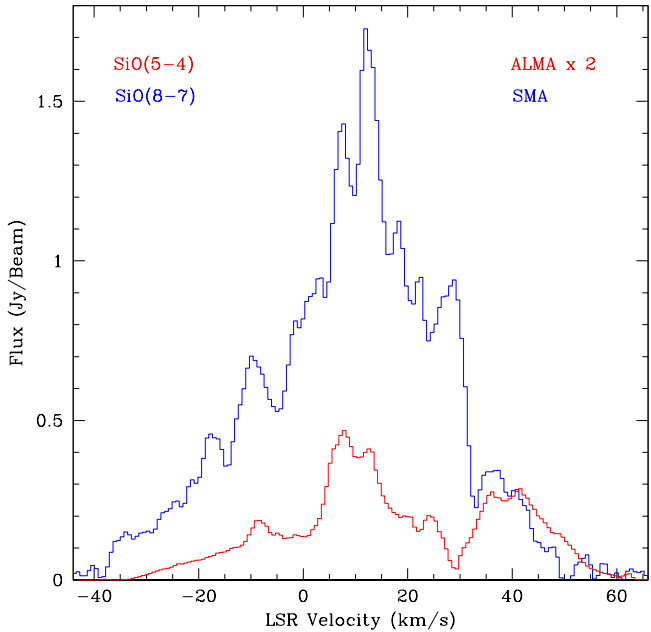}
\caption{\scriptsize SiO(5-4) (red line) and SiO(8-7) (blue line) spectra from the Orion BN/KL region obtained with 
                     the ALMA and SMA observations, respectively. The spectrum produced from the ALMA data 
                     is multiplied by a factor of two for convenience. In the ALMA data 1 Kelvin is equivalent to 9.04 Jy, while
                     in the SMA data 1 Kelvin is equivalent to 1.73 Jy.}
\label{fig1}
\end{center}
\end{figure}

The ALMA digital correlator was configured in 4 spectral windows of 1875 MHz and
3840 channels each.  This provides a spectral resolution of 0.488 MHz ($\sim$ 0.7 km s$^{-1}$) 
per channel. As a special case, this observation consisted of 5 different spectral tunings resulting
on 20 spectral windows. The spectral window where the line SiO (J=5-4; $\nu$=0) at a rest 
frequency of 217.10498 GHz was detected, has a central frequency of 217.225 GHz.

Observations of Callisto provided the absolute scale for the flux density calibration while observations of the
the quasar J0607$-$085 (with a flux of 1.4 Jy) provide the gain phase calibration.  
The total on-source integration time was about 100 min.  This time was divided to observe the 5 spectral tunings.
For time-dependent gain calibration, the nearby quasar J0607$-$085 was observed every 7 minutes. 

The data were calibrated, imaged, and analyzed using the Common Astronomy Software Applications (CASA).  
To analyze the data, we also used the KARMA software \citep{goo96}.  
The resulting r.m.s.\ noise for the line images was about 100 mJy beam$^{-1}$ at an angular resolution of $1\rlap.{''}78$ 
$\times$ $1\rlap.{''}23$ with a P.A. = $-4.4^\circ$. We used uniform weighting to obtain a slightly 
better angular resolution.   

\begin{figure*}[h!]
\begin{center}
\includegraphics[scale=0.85]{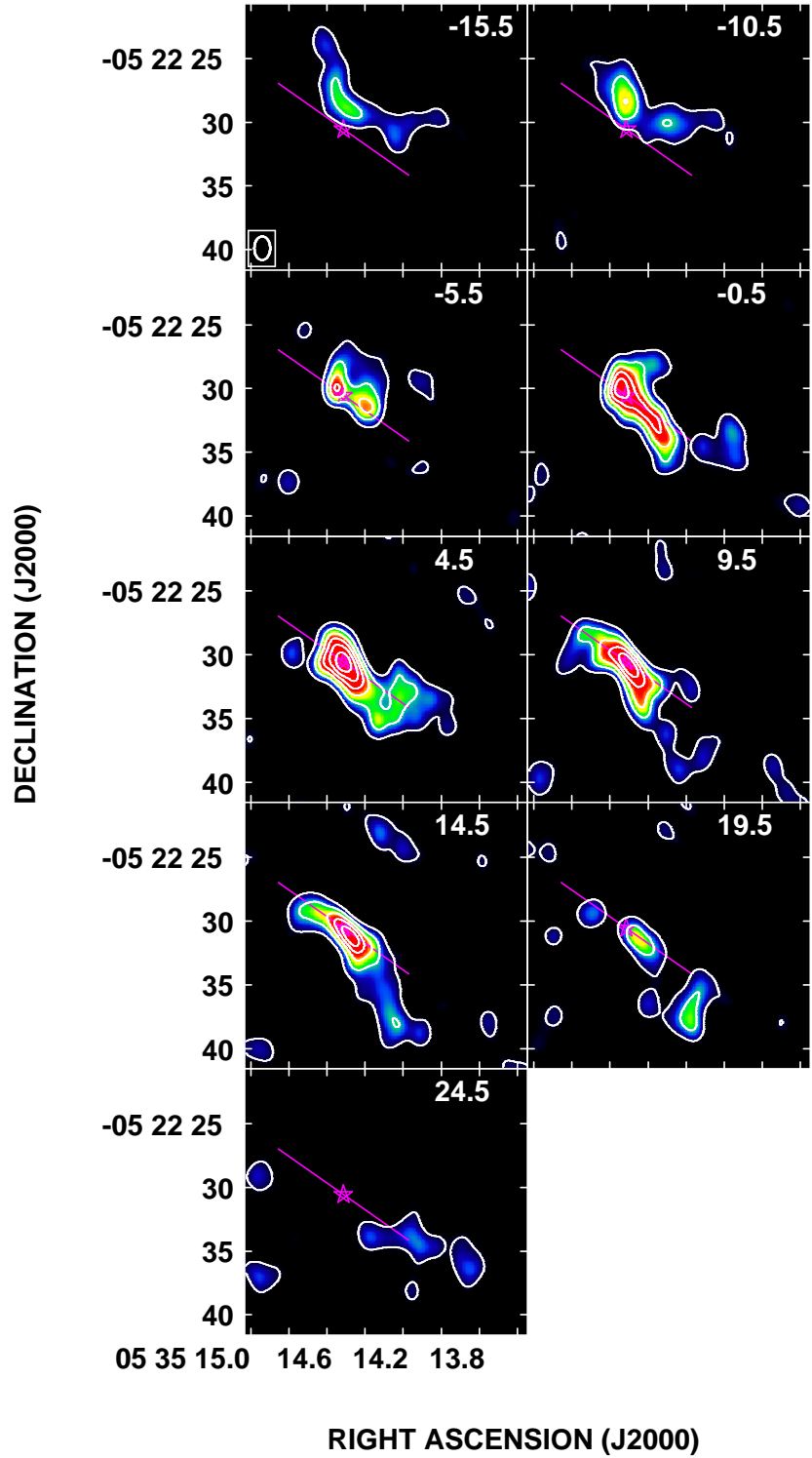}
\caption{\scriptsize ALMA channel map of the SiO (J=5-4; $\nu$=0) line toward the Orion Source I. Every velocity channel is 
5.0 km s$^{-1}$ wide and is centered close to the position of the Orion Source I. Each box is labeled with the 
central LSR velocity in km s$^{-1}$. A pink star marks the location of radio Orion Source I. The line represents the orientation of the outflow 
at a P.A.= 55$^\circ$ \citep{plam2009}. The contours are from -5, 5, 10, 15, 20, 25, 30, 35, 40, 45, 50, 55, and 
60 times 0.14 Jy beam$^{-1}$, the rms noise of the image. The half-power contour of the synthesized beam is 
shown in the bottom left corner of the first panel.}
\label{fig3}
\end{center}
\end{figure*}

\subsection{SMA}

The observations were obtained from the SMA, and were collected on December 2009, 
when the array was in its compact configuration.  The 21 independent baselines in the compact 
configuration ranged in projected length from 16 to 70 k$\lambda$.  The phase reference center 
for the observations was at $\alpha_{J2000.0}$ = \dechms{05}{35}{15}{45}, $\delta_{J2000.0}$ = \decdms{05}{22}{35}{0}.
Two frequency bands, centered at 335.3306310 GHz LSB (Lower Sideband) and 
345.3306310 GHz USB (Upper Sideband) were observed simultaneously. The observations were made in 
mosaicing mode using half-power point spacing between field centers and thus covering the 
entire BN/KL region. The primary beam of the SMA at 345 GHz has a FWHM $\sim 30''$. 

The SMA digital correlator was configured in 24 spectral windows (``chunks'') of 104 MHz and
128 channels each. This provides a spectral resolution of 0.812 MHz ($\sim$ 0.7 km s$^{-1}$) 
per channel. The zenith opacity ($\tau_{230 GHz}$), measured with the NRAO tipping 
radiometer located at the nearby Caltech Submillimeter Observatory, was about 0.07 
, indicating excellent weather conditions during the observations. Observations of 
Uranus provided the absolute scale for the flux density calibration.  The gain calibrators were the 
quasars J0541-056 and J0607-085, while 3C454.3 was used for bandpass calibration. The uncertainty in 
the flux scale is estimated to be between 15 and 20$\%$, based on the SMA monitoring of 
quasars.  Further technical descriptions of the SMA and its calibration schemes can be
found in \citet{Hoetal2004}.

The data were calibrated using the IDL superset MIR, originally developed for the Owens Valley 
Radio Observatory \citep[OVRO,][]{Scovilleetal1993} and adapted for the SMA.\footnote{The 
MIR-IDL cookbook by C. Qi can be found at http://cfa-www.harvard.edu/$\sim$cqi/mircook.html.} 
The calibrated data were imaged and analyzed in the standard manner using the MIRIAD and
KARMA \citep{goo96} softwares.  We set the ROBUST parameter of the task INVERT to 0 to obtain 
an optimal compromise between resolution and sensitivity. The r.m.s.\ noise in each channel of the 
spectral line data was about 170 mJy beam$^{-1}$ at an angular resolution of $2\rlap.{''}9$ 
$\times$ $1\rlap.{''}9$ with a P.A. = $+16.6 ^\circ$. The SiO J=8-7; $\nu$=0 was detected in the
USB at a rest frequency of about 347.3 GHz. 

\section{Results}

In Figure 1, we show the total power SiO spectra of the ALMA and SMA observations toward 
the central part of the BN/KL region. The total area from the spectra was obtained covers approximately
one arcmin$^2$ centered at the position of the Source I. No obvious maser emission is detected at these transitions.
The spectra are very similar, but with that obtained at higher frequencies 
(SMA) being much stronger than the obtained with ALMA. This has also been observed in outflows 
associated with young low mass stars  \citep[{\it e.g.} HH 211;][]{lee2007,Palau2006}
and is attributed to the SiO excitation within the outflow. The spectra have two central strong peaks,
one peaking around of $+$6 $\kms$ and the second approximately at $+$12 $\kms$. These peaks are associated with the emission from the
 outflow of the Orion Source I around the systemic velocity of the cloud ($+$9 $\kms$). 
 The dip at 9 $\kms$ could be self-absorption. The high-velocity structure
has multiple components, some of them come from the outflow of the Orion Source I, but others come from the molecular bubble
reported by \citet{zapata2011b}, and objects located in BN/KL.  
We estimated that the radial velocity of the Orion Source I is about $+$6 $\kms$, in good agreement with that value
 found by \citet{plam2009} and \citet{matt2010}.   

\begin{figure*}[ht]
\begin{center}
\includegraphics[scale=0.39]{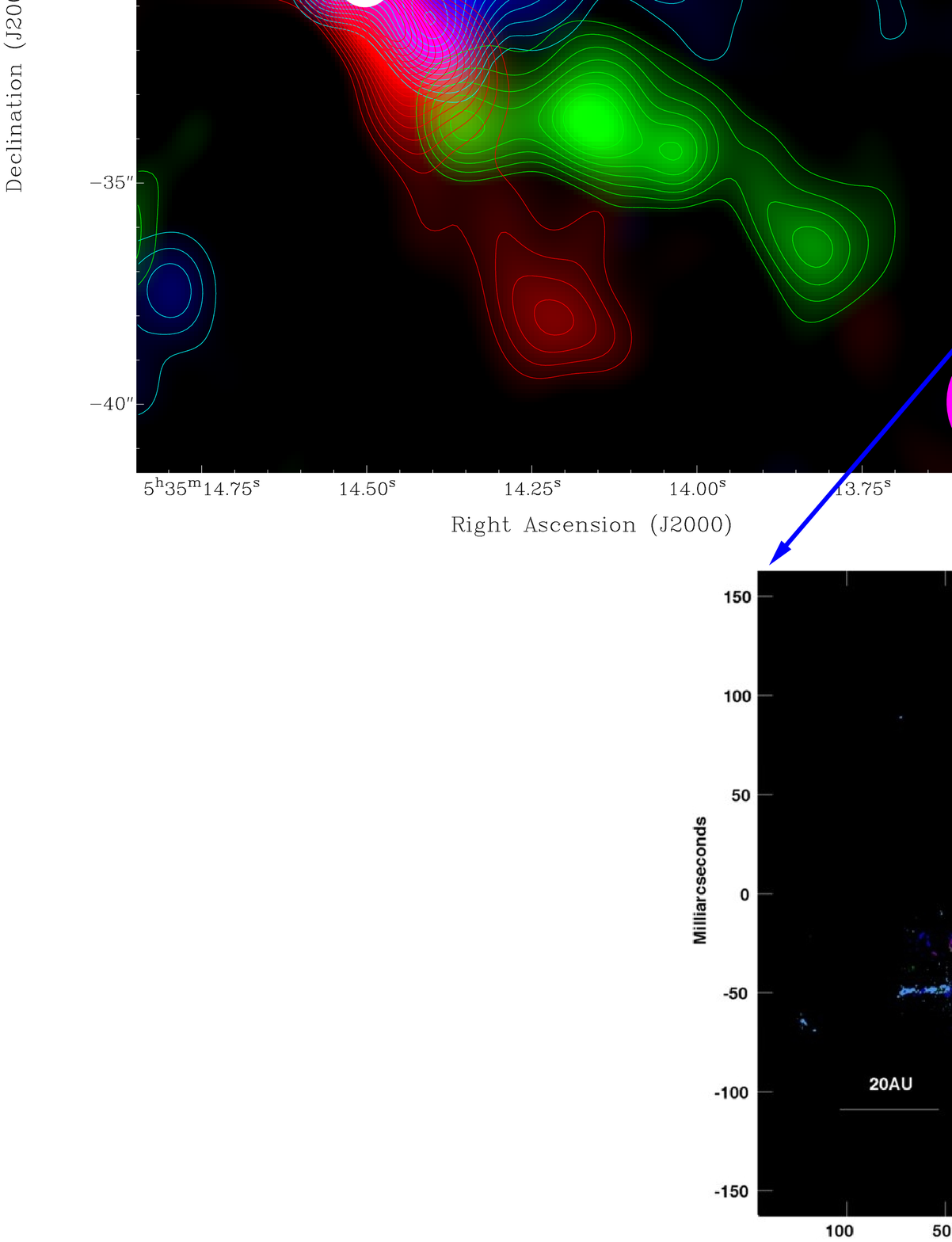}
\caption{\scriptsize {\it Upper panels:} ALMA (left) and SMA (right) integrated intensity (moment 0) color-scale and 
contour maps of the SiO thermal emission from the outflow arising from the Orion Source I. 
The blue color-scale  and contours represent blueshifted gas with LSR velocities ranking 
from $-$15 km s$^{-1}$ to $-$5 km s$^{-1}$. The red color-scale and contours represent redshifted 
gas with LSR velocities ranging from $+$15 
km s$^{-1}$ to $+$25 km s$^{-1}$.  The green color-scale and contours represent redshifted high velocity gas 
with LSR velocities ranging from $+$26 km s$^{-1}$ to $+$30 km s$^{-1}$. In the left panel, the
blue and red contours range from $-$10\% to 90\% of the peak emission, in steps of 5\%. 
The emission peaks for the blueshifted and redshifted emission are 28 and 59 Jy beam$^{-1}$ km s$^{-1}$. 
The green contours range from $-$25\% to 90\% of the peak emission, in steps of 10\%.
The emission peak for the redshifted emission (associated with the green color) is 27 Jy beam$^{-1}$ km s$^{-1}$.
In the right panel, the
blue and red contours range from $-$20\% to 90\% of the peak emission, in steps of 5\%. 
The emission peaks for the blueshifted and redshifted emission are 140 and 350 Jy beam$^{-1}$ km s$^{-1}$, respectively. 
The green contours range from 40\% to 90\% of the peak emission, in steps of 10\%.
The emission peak for the redshifted emission (associated with the green color) is 100 Jy beam$^{-1}$ km s$^{-1}$.
In both panels, the synthesized beams of the observations are shown in the lower right corner.
The white dot marks the position of the Orion Source I, as derived from VLA proper-motion data \citep{gom2008}. 
{\it Lower panel:} This image was obtained from \citet{matt2010} and displays the velocity 
field of the $^{28}$SiO  $v = 1$ and $v = 2$, J = 1-0 emission surrounding the Orion Source I.  }
\label{fig2}
\end{center}
\end{figure*}

The channel velocity map of the SiO (J=5-4; $\nu$=0) emission from the Orion Source I mapped with ALMA is shown in Figure 2. 
The range of radial velocities ($-$15.5 to $+$24.5  $\kms$) is in good agreement with the one obtained 
by \citet{plam2009} shown in their Figure 1. However, our maps with lower angular resolution seem to recover
some extended parts of the outflow lost in their maps. This is more evident in our channels maps with
central velocities of $-$15.5 $\kms$ and $+$14.5 $\kms$. In these channels, we observe some north-south
structures with a \textquotedblleft wing'' shape, extending in an east-west orientation. The emission arising from the central velocity 
channels could not be resolved with our actual resolution, as in \citet{plam2009}.  

Integrated intensity maps (moment 0) of the SiO emission obtained with ALMA and SMA are presented in Figure 3.
In these maps, we only integrated over redshifted and blueshifted radial velocities associated
with the outflow from the Orion Source I. The blueshifted gas emission was integrated 
over LSR velocities ranging from $-$15 km s$^{-1}$ to $-$5 km s$^{-1}$. 
The redshifted gas emission was integrated over the LSR velocities range
from $+$15 km s$^{-1}$ to $+$25 km s$^{-1}$. The very high redshifted velocities range 
from $+$26 km s$^{-1}$ to $+$30 km s$^{-1}$. These maps reveal that the outflow from the Orion Source I has a marked \textquotedblleft butterfly''
 morphology along a northeast-southwest axis, with the blueshifted radial velocities found to the northwest, 
while the redshifted velocities are toward the southeast. This velocity structure is different of what is 
observed at smaller scales, see the Figure 7 of \citet{matt2010} or our Figure 3. The ranges of radial velocities of the SiO masers
 are similar to those obtained by us. The wings of the \textquotedblleft butterfly'' shape extend several arcseconds 
 along the east-west direction.   

\begin{figure}[ht]
\begin{center}
\includegraphics[scale=0.28]{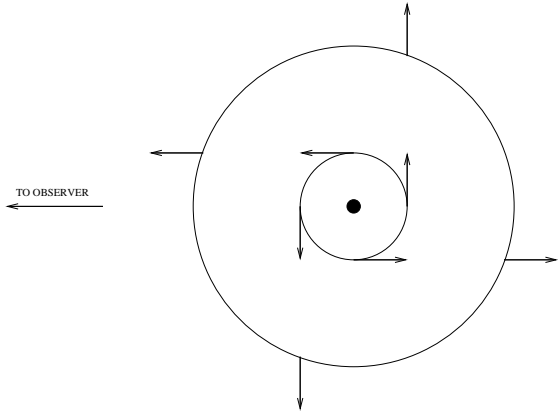}
\caption{\scriptsize The central dot marks the position of the star and the inner circle indicates a rotating disk. 
Gas ejected ballistically from the disk will appear, at larger scales, radially expanding (and not rotating).}
\label{fig4}
\end{center}
\end{figure}

At the middle of the \textquotedblleft wings'', we find very high velocity emission (see Figure 2, the green color-scale and contours) 
that also appears to emanate from the Source I, this
component is observed at small scales with the same orientation by \citet{matt2010}. In the ALMA map, with a better sensitivity
and angular resolution than the SMA one, the tail extends a few more arcseconds. The \textquotedblleft butterfly'' morphology can also be
seen in the 0.5$''$ resolution map of \citet{wri1995} in the SiO (J=2-1; $\nu$=0) and in Figure 3
of \citet{plam2009} which was derived from 1$'$ resolution data .  The wings of the butterfly also show some differences between both maps. The red wing is only half, different from the blue wing that is shaped like an U in the ALMA map. Something seems to suppress the other red half.  Maybe it is due to some density inhomogeneities in the surroundings of the bipolar outflow.
    
\section{Discusion}

In the following, we will discuss some possibilities trying to explain the nature of the velocity structure and 
morphology of the outflow emanating from Source I.
  
\subsection{Two bipolar outflows}

A likely possibility is to have two bipolar outflows emanating from a short-period binary star located 
in the middle of the ionized disk of the Source I. This is in agreement with what was proposed by 
\citet{bally2011} about the binarity of the Source I. One outflow
could be moving in a nearly north-south orientation with its blueshifted side in the north and the
redshifted side toward the south. The second bipolar outflow, would have an east-west axis, with its blueshifted
side to the west, while its redshifted side is to the east. The difference in size of the components could 
be attributed to the molecular cloud inhomogeneities. The fact that a binary could have a strong precession moving
in short time scales, might explain the difference in the redshifted and blueshifted velocities observed
at different scales. The SiO masers could be tracing the innermost parts of the two bipolar outflows ejected 
at the present time \citep{matt2010}. However, this hypothesis could not explain satisfactorily well the high velocity tail 
(with a southwest orientation) observed in the middle of the outflows, and the observations of the single ''cone'' 
outflow found by \citet{plam2009} at their central channel velocities.

\subsection{Ballistic ejecta from a rotating disk}

We also consider the possibility that the thermal SiO traces gas that was ejected ballistically in the past 
from the small-scale disk associated with Source I. However, Figure 4 shows that the gas ejected 
ballistically at small scales from a rotating structure will, at large scales, appears as an expanding structure,
with the largest radial velocities overlapping at the center of the structure. This is not what is observed and we rule out
this possibility.     

\subsection{A single expanding wide-angle NE-SW outflow}

Another possibility is having a wide-angle expanding outflow that is emanating from the ionized, nearly edge-on, rotating disk 
reported by \citet{matt2010} and \citet{men2007}. 
In this case, the SiO masers would trace the innermost parts of the outflow and the thermal SiO emission would be tracing the outer parts of the outflow.
The difference in the orientation of the radial velocities 
at different scales could be attributed to that maybe in the past the rotation of the disk was in a different sense. 
This could have happened if a large amount of matter fell directly to the disk and changed the rotation sense. 
This third idea might explain most of the characteristics observed in Figure 3 and those observed at small scales
by \citet{matt2010} and \citet{plam2009}. However, the idea of a disk inverting its rotation direction requires of unusual 
events during the evolution of the disk and deserves further theoretical and observational studies. 

\section{Conclusions}

In this paper, we have reported SiO (J=5-4; $\nu$=0) and SiO (J=8-7; $\nu$=0) 
observations obtained with ALMA and the SMA. 
We found a marked \textquotedblleft butterfly'' morphology along a northeast-southwest axis in the outflow from the Source I. 
However, the blueshifted radial velocities are found to the northwest, 
while the redshifted velocities toward the southeast, 
which is contrary to what is observed at very small scales. 
We discussed some possibilities to explain this phenomenon, however, 
any of these hypotheses requires further sensitive, and high spatial resolution observations. 

\acknowledgments
This paper makes use of the following ALMA Science Verification data: ADS/JAO.ALMA\#2011.0.00009.SV 
L.A.Z, L. L. and L. F. R. acknowledge the financial support from DGAPA, UNAM, and CONACyT, M\'exico. 
L. L. is indebted to the Alexander von Humboldt Stiftung for financial support. We are very
thankful for the suggestions of anonymous referee that helps to improve our manuscript.


\begin{thebibliography}{}
\bibitem[Bally et al.(2011)]{bally2011} Bally, J., Cunningham, 
N.~J., Moeckel, N., et al.\ 2011, \apj, 727, 113 
\bibitem[Churchwell et al.(1987)]{chu1987} Churchwell, E., 
Felli, M., Wood, D.~O.~S., \& Massi, M.\ 1987, \apj, 321, 516 
\bibitem[Ho et al.(2004)]{Hoetal2004} Ho, P.~T.~P., Moran, J.~M., \& Lo, K.~Y.\ 2004, \apjl, 616, L1 
\bibitem[Genzel et al.(1981)]{gen1981} Genzel, R., Reid, M.~J., 
Moran, J.~M., \& Downes, D.\ 1981, \apj, 244, 884
\bibitem[Goddi et al.(2011)]{god2011} Goddi, C., Humphreys, 
E.~M.~L., Greenhill, L.~J., Chandler, C.~J., 
\& Matthews, L.~D.\ 2011, \apj, 728, 15 
\bibitem[Gooch(1996)]{goo96} Gooch, R.\ 1996, Astronomical Data Analysis Software and Systems V, 101, 80 
\bibitem[G{\'o}mez et al.(2008)]{gom2008} G{\'o}mez, L., 
Rodr{\'{\i}}guez, L.~F., Loinard, L., et al.\ 2008, \apj, 685, 333 
\bibitem[Lee et al.(2007)]{lee2007} Lee, C.-F., Ho, P.~T.~P., 
Palau, A., et al.\ 2007, \apj, 670, 1188 
\bibitem[Matthews et al.(2010)]{matt2010} Matthews, L.~D., 
Greenhill, L.~J., Goddi, C., et al.\ 2010, \apj, 708, 80
\bibitem[Menten 
\& Reid(1995)]{men1995} Menten, K.~M., \& Reid, M.~J.\ 1995, \apjl, 445, L157
\bibitem[Palau et al.(2006)]{Palau2006} Palau, A., Ho, P.~T.~P., 
Zhang, Q., et al.\ 2006, \apjl, 636, L137
\bibitem[Plambeck et al.(2009)]{plam2009} Plambeck, R.~L., 
Wright, M.~C.~H., Friedel, D.~N., et al.\ 2009, \apjl, 704, L25
\bibitem[Scoville et al.(1993)]{Scovilleetal1993} Scoville, N.~Z., Carlstrom, J.~E., Chandler, C.~J., et al.\ 1993, \pasp, 105, 1482
\bibitem[Reid et al.(2007)]{men2007} Reid, M.~J., Menten, 
K.~M., Greenhill, L.~J., \& Chandler, C.~J.\ 2007, \apj, 664, 950
\bibitem[Rodr\'\i guez et al.(2005)]{rod2005}Rodr{\'{\i}}guez, L.~F., Poveda, A., Lizano, S., 
\& Allen, C.\ 2005, \apjl, 627, L65 
\bibitem[Wright et al.(1995)]{wri1995} Wright, M.~C.~H., 
Plambeck, R.~L., Mundy, L.~G., \& Looney, L.~W.\ 1995, \apjl, 455, L185 
 \bibitem[Zapata et al.(2009)]{zap2009} Zapata, L.~A., 
Schmid-Burgk, J., Ho, P.~T.~P., Rodr{\'{\i}}guez, L.~F., 
\& Menten, K.~M.\ 2009, \apjl, 704, L45 
\bibitem[Zapata et al.(2011a)]{zapata2011a} Zapata, L.~A., Schmid-Burgk, J., \& Menten, K.~M.\ 2011b, \aap, 529, A24 
\bibitem[Zapata et al.(2011b)]{zapata2011b} Zapata, L.~A., Loinard, L., Schmid-Burgk, J., et al.\ 2011a, \apjl, 726, L12 
\end{thebibliography}
\end{document}